\documentclass[10pt]{revtex4}
\usepackage{amsmath,amssymb}
\usepackage{graphicx,epstopdf}
\usepackage[usenames,dvipsnames,svgnames]{xcolor}
\usepackage{hyperref}
\usepackage{array}
\usepackage{multirow}
\begin{document}

\title{Static spherically symmetric wormhole with quadratic shape function and particle trajectories around it}
\author{Bikram Ghosh$^1$\footnote {bikramghosh13@gmail.com}}
\author{Sourav Dutta$^2$\footnote {sduttaju@gmail.com}}
\author{Sudeshna Mukerji$^3$\footnote {mukerjisudeshna@gmail.com}}
\author{Subenoy Chakraborty$^3$\footnote {schakraborty.math@gmail.com}}
\affiliation{$^1$Department of Mathematics, Ramakrishna Mission Vidyamandira, Howrah-711202, West Bengal, India\\
	$^2$ Department of Mathematics, Dr. Meghnad Saha College, Uttar Dinajpur-733128, West Bengal, India\\
	$^3$ Department of Mathematics, Jadavpur University, Kolkata-700032, West Bengal, India}

	
\begin{abstract} The paper deals with static traversable wormhole having shape function a polynomial of the radial coordinate of degree 2. Embedding of the wormholes in space-time is examined both analytically and graphically. Also both timelike and null geodesics are studied in this wormhole geometry.
\end{abstract}
\maketitle

\section{Introduction:}
Wormholes in general relativity are the solutions of the Einstein Field Equations(EFE) and are act as hypothetical tunnels connecting \cite{r1} two asymptotically flat universes or two asymptotically flat portions of the same universe. They can be considered as one of the most interesting predictions of general relativity. Also wormholes are considered as shortcuts connecting distant regions of spacetime --- the notion of time machine \cite{r2.1}. The space-time geometry of wormholes are topologically non simply connected. The idea of wormhole was initiated by Flamm \cite{r2} immediatlely, after Einstein's formulation of gravity  and subsequently by Einstein and Rosen \cite{r3} and Wheeler \cite{r4}. Infact the name wormhole was first coined by John Wheeler {\cite{r4}-\cite{r5}} in 1962 by reinterpreting the Einstein-Rosen bridge \cite{r3} as a connector between two distant places in spacetime having no mutual interaction. The modern development of wormhole geometry was initiated by Morris and Thorne \cite{r6} by introducing the idea of traversability {\cite{r7}-\cite{r8.1}} which gives the possibility for a human being to travel unchanged through the tunnel of the wormhole in both directions in a finite time. As a consequence, there should not be any spacetime singularity \cite{r9} or horizon while there induces bearable tidal force \cite{r10}, so for a fascinating application one may dream of travelling between distanct galaxies. The common way of formulating wormhole solution is to construct singularity free spacetime metric and the matter contained is determined from the EFEs. Usually, this matter contained in wormhole spacetime does not obey the weak energy condition (exotic matter \cite{r10.1}). There are lot of attempts to minimize the violation of the energy conditions either by considering arbitrary small exotic matter or the exotic matter is restricted only to particular regions {\cite{r11}--\cite{r13}}. For an elaborate discussion on energy conditions see the work of Visser {\cite{r1},\cite{r11}}.\par
However, at present such type of exotic matter fields are seem to be favourable due to observed accelerated expansion of the universe. It is speculated that the unusual matter field is nothing but the gravitational effects of the dark energy or phantom energy, violating the energy conditions. Subsequently, static spherical wormholes have been constructed \cite{r14} with phantom matter field in the form of anisotropic fluid having large negative radial pressure so that $\omega_r=\frac{p_r}{\rho}<-1.$ \par
The present work deals with such traversable wormhole with a quadratic shape function, and the construction approach is along the line of the conventional way of Morris and Thorne \cite{r6}.
The paper is organized as follows: Traversable wormhole with quadratic shape function has been described in section \ref{sec:ii}. Section \ref{sec:iii} deals with wormhole space-time geometry and embedding. Geodesics in wormhole spacetime has been presented in section \ref{sec:iv} by Hamilton-Jacobi approach. In section \ref{sec:V}, phantom trajectories are examined for possibility of shadow of wormhole. Section \ref{sec:vi} deals with time-like geodesics with bounded and unbounded orbits. The paper ends with a brief discussion in section \ref{sec:vii}. 
\section{ Traversable wormhole with quadratic shape function}
\label{sec:ii}
The two familiar ways in which static spherically symmetric wormhole metric can be described as \cite{r6}
\par 
$(i)$ radial co-ordinate system:
 \begin{equation}\label{eq1}
ds^2=e^{2\phi(r)}dt^2-\frac{dr^2}{1-\frac{b(r)}{r}}-r^2d\Omega_2^2,   
\end{equation}
\par
 $(ii)$ proper radial co-ordinate system:
 \begin{equation}
  dS^2=e^{2\Phi(R)}dt^2-dR^2-r^2(R)d{\Omega_2}^2,    
\end{equation}
where `$t$' is global time, the shape function $b(r)$ is a function of the radial co-ordinate, $\phi(r)$(\text{\it i.e., }$\Phi(R))$ is the redshift function, $d\Omega_2^2=d\theta^2+{\sin^2\theta}d\phi^2$ is the metric on unit two sphere and $R$, the proper radial co-ordinate is given by  $R=\int_{r_0}^{r}\frac{dr}{\sqrt{1-\frac{b(r)}{r}}}$, $r_0$ being the throat radius. Note that the surface area of 2-hypersurface: $t=$ constant, $R=$ constant is $4\pi r^2$. The proper radius $R$ extends over the whole space-time while the radial co-ordinate `$r$' ranges $(r_0, \infty)$ so that one needs two charts to cover the whole space-time. Further, the common interesting region of the two distant regions $R\rightarrow$$\pm\infty$ is known as the throat of the wormhole. Hence `$r$' behaves as Schwarzschild-like radial co-ordinate. 
\par Now for the existence of wormhole, the shape function $b(r)$ is restricted \cite{r1},\cite{r6},\cite{r7} as follows:\\
$	(i)$
\begin{equation}
\label{eq5}
 b(r_0)=r_0,
\end{equation} 
where $r_0$ is the location of the throat having minimum radius. Thus radial coordinate ranges over $r_0\leq r<\infty.$\\
$(ii)$
\begin{equation}\label{eq6}
\frac{b(r)}{r}\leq 1
\end{equation}
to keep the metric to be Lorentzian in nature. The equality sign will occur at the throat.\\
$(iii)$
\begin{equation}\label{eq7}
\frac{b(r)}{r}\rightarrow 0 ~~as~~ r\rightarrow \infty
\end{equation}
is the requirement for asymptotic flatness. This restriction leads to a geometry connecting two asymptotically flat regions.\\
$(iv)$
\begin{equation}\label{eq8}
\frac{b-b^\prime r}{2b^2}>0
\end{equation}
is the condition for flare out \cite{r14.1}. It gives the geometric condition for the minimality of the wormhole throat. Physically, it means the divergence of the null rays through throat{ \it i.e.,} violation of null energy condition. Moreover, the redshift function $\phi(r)$ should be non-zero and finite through the spacetime to avoid the presence of horizons and singularities.\par
In the present work the shape function is chosen as quadratic function of `$r$'
 with two independent coefficient due to the throat condition. Depending on the sign choices of these two coefficient four possible types of solutions are possible of which two will give wormhole solution extending to infinity while third one is a wormhole geometry restricted to a finite region and the fourth one is a purely spherically symmetric static solution (as flare-out condition is violated). The quadratic form of the shape function in wormhole geometry was first used in \cite{r20}. The reason for choosing the quadratic shape function is that two types of wormhole geometry with exotic matter is possible for this choice of the shape function namely: $(i)$ infinitely extended phantom wormhole and $(ii)$ finitely extended phantom wormhole connected to inhomogeneous and anisotropic spherically symmetric distribution of dark energy. Further, the exotic matter is very much important in the context of present accelerated expansion. In the present work the quadratic form of shape function is chosen as
\begin{equation}
b(r)=\alpha r^2+\beta r+\delta
\end{equation}
where the constant parameters $\alpha$, $\beta$, $\delta$ are either restricted or determined from the above wormhole conditions (\ref{eq5})---(\ref{eq8}) as follows:\\
Equation (\ref{eq5}) determines $\beta$ of the form
\begin{equation}\label{eq10}
\beta=1-\alpha r_0-\frac{\delta}{r_0},
\end{equation}
so that 
\begin{equation}\label{eq11}
b(r)=r-r\left(1-\frac{r_0}{r}\right)\left(\frac{\delta}{r_0}-\alpha r\right).
\end{equation}
Condition (\ref{eq6}) gives
\begin{equation}
\left(1-\frac{r_0}{r}\right)\left(\frac{\delta}{r_0}-\alpha r\right)>0.
\end{equation}
As $r>r_0$, so one has 
\begin{equation}\label{eq13.1}
\frac{\delta}{r_0}-\alpha r>0.
\end{equation}
Also the flare-out condition (\ref{eq8}) gives the restriction
\begin{equation}\label{eq14.1}
\delta-\alpha r^2>0.
\end{equation}
The above two inequalities (\ref{eq13.1}) and (\ref{eq14.1}) show that among the four possible sign choices for $\alpha$ and $\delta$, the choice ($\alpha>0,\delta<0$) does not satisfy them ({\it i.e.,} it gives spherically symmetric solution) and one has three possible choices for wormhole geometry presented in the following table: \begin{table}[!htb]
	\centering
	\caption{Sign choice of $\alpha$ and $\delta$, and wormhole geometry}
	\begin{tabular}{|>{\bfseries}c|*{5}{c|}}\hline
		{\bfseries Sign of $\alpha$ and $\delta$} & {\bfseries Range of  $r$} & {\bfseries $g^{rr}=1-\frac{b(r)}{r}$}
		 \\ \hline
		\text{WH 1: $\alpha>0$, $\delta>0$}         & \text{Finite size: $r_0<r<\sqrt{\frac{\delta}{\alpha}}$} & \text{ $\frac{\alpha}{r}(r-r_0)(\mu_0 r_0-r)$}      \\  
		&\text{ with $r_0<\sqrt{\frac{\delta}{\alpha}}$} & \text{with $\sqrt{\frac{\delta}{\alpha}}=\mu_0 r_0$, $\mu_0>1$} \\ \hline
		\text{WH 2: $\alpha<0$, $\delta>0$} & \text{Infinitely extended:} & \text{$\frac{|\alpha|}{r}(r^2-r_0^2)$} \\
		&$r_0<r<\infty$ & \text{choosing $r_0^2=\frac{\delta}{|\alpha|}$}
		 \\
		\hline
		\text{WH 3: $\alpha<0$, $\delta<0$}& \text{Infinitely extended:} & \text{$\frac{|\alpha|}{r}(r-r_0)^2$}\\
		& \text{$r_0=\sqrt{|\frac{\delta}{\alpha}|}<r<\infty$} &
		 \\ 
		\hline
		
	\end{tabular}
\label{Table:T1}
\end{table}\\\\
%
\\
Note that although WH 2 and WH 3 are extended to infinity but none of them is asymptotically flat and hence they are not of Morris and Thorne type.
\par
Now considering anisotropic fluid with energy-momentum tensor
$$T_\mu^\nu=\text{diag}(\rho, -p_r, -p_t, -p_t),$$
the Einstein field equations for the metric (\ref{eq1}) can be written as \cite{r14.0}
\begin{equation}\label{eq2}
\frac{b^\prime}{r^2}=\kappa\rho(r),
\end{equation}
\begin{equation}\label{eq3}
2\left(1-\frac{b}{r}\right)\frac{\phi^\prime(r)}{r}-\frac{b}{r^3}=\kappa p_{r}(r),
\end{equation}
\begin{equation}\label{eq4}
\left(1-\frac{b}{r}\right)\big[{\phi^{\prime\prime}}+({\phi^\prime})^2-\frac{b^\prime r+b-2r}{2r(r-b)}{\phi^\prime}-\frac{(b^\prime r-b)}{2r^2(r-b)}\big]=\kappa p_t(r)
\end{equation}
where as usual $\rho$, $p_r$ and $p_t$ are respectively the energy density, radial and transverse thermodynamical pressures of the fluid and an over prime denotes differentiation with respect to the radial coordinate `$r$'. 
\par
Now it is interesting to examine the energy conditions in the space-time geometry of the above wormhole configurations. A necessary condition for the existence of a static wormhole is the violation of the dominant energy condition (DEC):
$$ \rho \geq0,~~ \rho+p_r\geq0,~~ \rho+p_t\geq0 .$$
However, physically one should have positivity of the energy density everywhere. Further if $\phi(r)=$constant is chosen for simplicity then the relation $\rho+p_r+2p_t=0$ holds everywhere and hence the strong energy condition:$\rho+p_{\text{total}}\geq0$ is automatically satisfied. From the field equations (\ref{eq2})---(\ref{eq4}) with $\phi(r)=0$ and $b(r)$ from equation (\ref{eq11}) the expressions for the energy density, pressure components are
\begin{eqnarray}
\rho(r)&=&\frac{\alpha r_0(2r-r_0)+r_0-\delta}{r_0r^2},\label{eq15}\\
p_r&=&-\frac{(r-r_0)(\alpha r-\frac{\delta}{r_0})+r}{r^3},\label{eq16}\\
p_t&=&\frac{\delta -\alpha r^2}{2r^3} .\label{eq17}
\end{eqnarray}
Also one has
\begin{eqnarray}
\rho+p_r&=&\frac{\alpha r^2-\delta}{r^3},\label{eq18}\\
\rho+p_t&=&\frac{3\alpha}{2r}+\frac{\delta}{2r^3}-\frac{\alpha+\alpha r_0^2-r_0}{r^2r_0} \label{eq19}
\end{eqnarray}
and
\begin{equation}
\rho+p_r+2p_t=0.\label{eq20}
\end{equation}\par
So at the throat the above expressions become 
\begin{equation}
\rho(r_0)=\frac{(\alpha r_0^2+r_0-\delta)}{r_0^3},
\end{equation}
\begin{equation}
\rho+p_r\big|_{r=r_0}=\frac{\alpha r_0^2-\delta}{r_0^3},
\end{equation} and
\begin{equation}
\rho+p_t\big|_{r=r_0}=\frac{(\alpha r_0^2+2r_0-\delta)}{2r_0^3}.
\end{equation}
The restrictions for the validity of the energy conditions for the above three types of wormholes are present in the following three tables:

 \begin{table}[!htb]
	\centering
	\caption{Energy conditions for WH 1}
	\begin{tabular}{|>{\bfseries}c|*{5}{c|}}\hline
		\multirow{2}{*}{\bfseries Condition} & \multicolumn{3}{c|}{\bfseries sign of} 
		 \\ \cline{2-4}
		& \textbf{$\rho$} & \textbf{$\rho+p_r$} & \text{$\rho+p_t$} \\ \hline
		\text{$r_0>\alpha r_0^2+\delta$}         & \text{$\geq0~\forall r$}       & \text{$<0,~\forall r$}       & \text{$>0~\forall r$}             \\ \hline
		\text{$\delta-\alpha r_0^2<r_0<\delta+\alpha r_0^2$} & \text{$\geq0~\forall r$}    & \text{$<0,~\forall r$}       & \text{$>0~\text{for }r<m~ \text{or}~ r>n,$}        \\
		 & & & \text{ $<0~\text{for} ~m<r<n~ \text{and} ~r_0<m$}    \\
		 \hline
		\text{$r_0<\delta-\alpha r_0^2$}   & \text{$>0~  \text{for}~r>r_2=\frac{D^2}{2\alpha r_0}>r_0$}    & \text{$<0,~\forall r$}   & \text{$>0~\text{for }r<m~ \text{or}~ r>n,$}        \\
		& & & \text{ $<0~\text{for} ~m<r<n$}               \\ \hline
	\end{tabular}
\label{Table:T2}
\end{table}
where $n,m=\frac{D^2}{3\alpha r_0}\Bigg[1\pm\sqrt{1-\frac{9\alpha r_0^2\delta}{D^4}}\Bigg],~D^2=\delta+\alpha r_0^2-r_0>0$.

\begin{figure}[!htb]
 	\centering
 	\begin{minipage}{.5\textwidth}
 		\centering
 		\includegraphics[width=.6\linewidth]{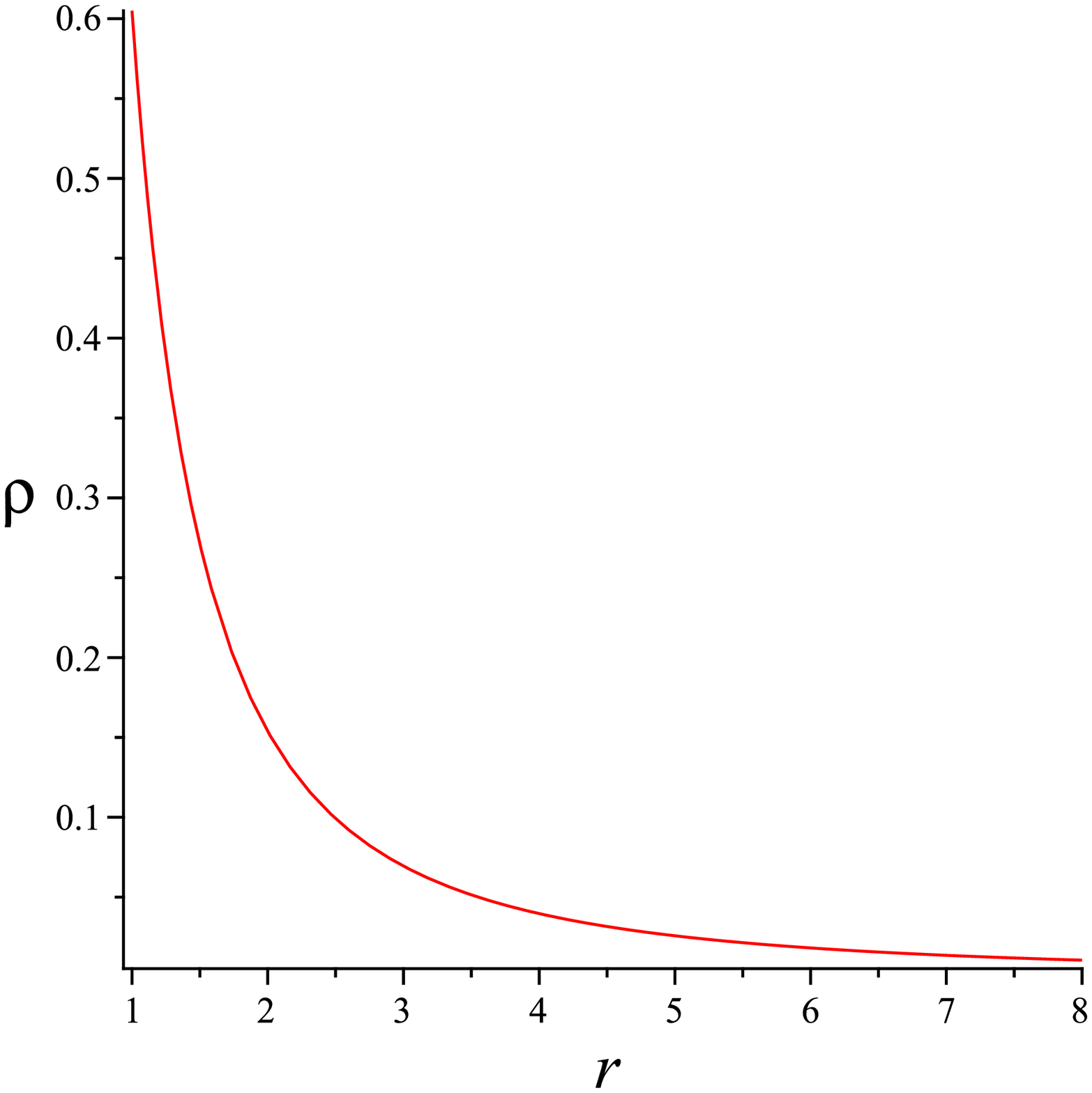}
 		\caption{Energy density curve ($\rho$) versus radial co-ordinate $r$\\ for WH 1 type ($\alpha=0.005$, $\delta=0.4$, $r_0=1$).}
 		\label{fig:f1}
 	\end{minipage}
 	\begin{minipage}{.5\textwidth}
 		\centering
 		\includegraphics[width=.6\linewidth]{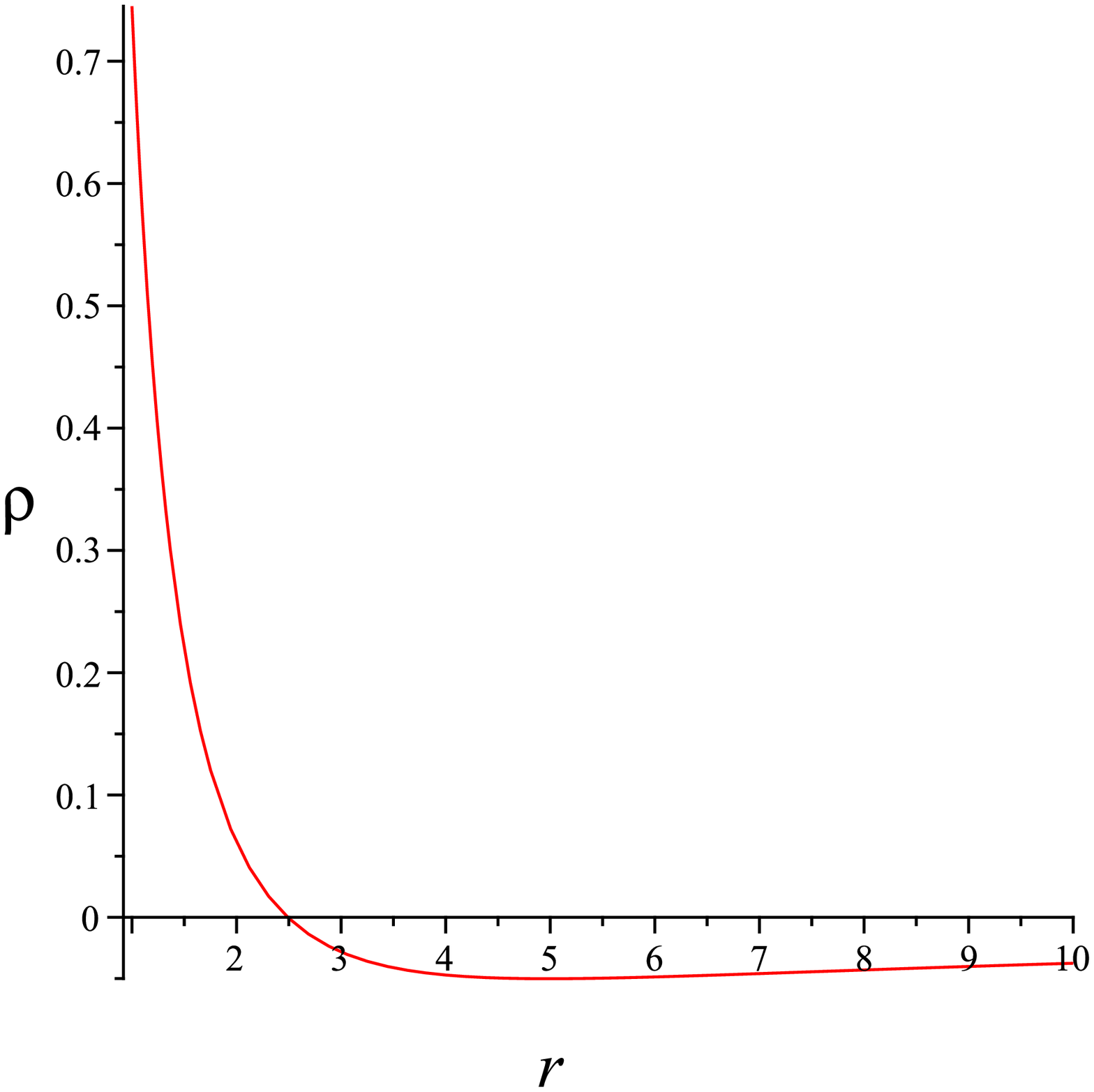}
 		\caption{Energy density curve ($\rho$) versus radial co-ordinate $r$\\ for WH 2 type ($\alpha=-0.25$, $\delta=0.005$, $r_0=1$).}
 		\label{fig:f2}
 	\end{minipage}
 \end{figure}

\begin{figure}[!htb]
	\includegraphics[width=.4 \textwidth]{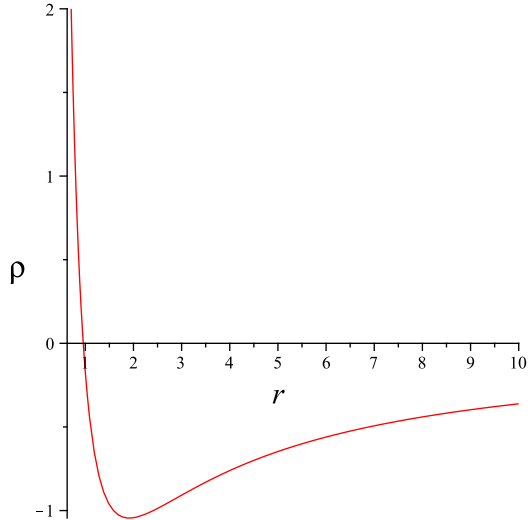}
	\caption{Energy density curve ($\rho$) versus radial co-ordinate $r$ for WH 3 type ($\alpha=-2$, $\delta=-1$, $r_0=1/\sqrt{2}$).}
	\label{fig:f3}
\end{figure}

 \begin{table}[!htb]
 	\centering
 	\caption{Energy conditions for WH 2}
 	\begin{tabular}{|>{\bfseries}c|*{5}{c|}}\hline
 		\multirow{2}{*}{\bfseries Condition} & \multicolumn{3}{c|}{\bfseries sign of} 
 		\\ \cline{2-4}
 		& \textbf{$\rho$} & \textbf{$\rho+p_r$} & \text{$\rho+p_t$} \\ \hline
 		\text{$r_0<r<r_1~ \text{where}~r_1=\frac{A}{2\left|\alpha\right|r_0},$} & \text{$>0$}    & \text{$<0,~\forall r$}       & \text{$~>0~\text{for }r<m_1,~ m_1=\sqrt{s+l^2}+l,$}        \\
 		\text{$A=r_0-\delta+\left|\alpha\right|r_0^2>0$}& & & \text{ $l=\frac{A}{3\left|\alpha\right|r_0},~s=\frac{\delta}{3\left|\alpha\right|}$}    \\
 		\hline
 		\text{$A<0$}   & \text{$~~~<0~~~$}    & \text{$<0,~\forall r$}   & \text{$~<0~\text{for}~r>m_1$}        \\
 		& & &                \\ \hline
 		
 	\end{tabular}
 \label{Table:T3}
 \end{table}
 \begin{table}[!htb]
 	\centering
 	\caption{Energy conditions for WH 3}
 	\begin{tabular}{|>{\bfseries}c|*{5}{c|}}\hline
 		\multirow{2}{*}{\bfseries Condition} & \multicolumn{3}{c|}{\bfseries sign of} 
 		\\ \cline{2-4}
 		& \textbf{$\rho$} & \textbf{$\rho+p_r$} & \text{$\rho+p_t$} \\ \hline
 		\text{$r_0<r<r_2~ \text{where}~r_2=\frac{\Lambda^2}{2\left|\alpha\right|r_0},$} & \text{$>0$}    & \text{$~>0~\text{if}~r<r_0$}       & \text{$~>0~\text{if }n_2<r<n_1~\text{or}~r_0<r<n_1,$}        \\
 		\text{$\Lambda^2=r_0+\left|\delta\right|+\left|\alpha\right|r_0^2$}& & & \text{ $\text{provided}~\left|\alpha\right|r_0^2<\left|\delta\right|+2r_0$}    \\
 		\hline
 		\text{$~r>r_2~$}   & \text{$~~~<0~~~$}    & \text{$<0,~\text{if}~r>r_0$}   & \text{$~<0~\text{if}~r>n_1~\text{and}~r_0<r<n_2$}        \\
 		& & & \text{provided $\left|\alpha\right|r_0^2>\left|\delta\right|+2r_0$}               \\ \hline
 		
 	\end{tabular}
 \label{Table:T4}\\
 \par 
    ~~~~~~~~~~~~~~~~~~~~~~~~~~~~~~~~~~~~~~~~~~~~~~~~~~~~~~~~~~~~~~~~~~~~~~~~~~~~~~~~~~~~~~~~~~~~~~~~~~~~~~~~~~~~~~~~~~~~~~~~~~~~~~~~~~~~~~~~~~~~~~~~~~~~~~~~~~~~~~~~~~~~~~~~~~~~~~~~~~~~~~~~~~~~~~~~~~~~~~~~~~~~~~~~~~~~~~~~~~~~~~~~~~~~~~~~~~~~~~~~~~~~~~~~~~~~~~~~~~~~~~~~~~~~~~~~~~~~~~~~~~~~~~~~~~~~~~~~~~~~~~~~~~~~~~~~~~~~~~~~~~~~~~~~~~~~~~~~~~~~~~~~~~~~~~~~~~~~
 \text{,} $\text{where}~~ n_1,n_2=l\pm\sqrt{l^2-s}$, $l=\frac{r_0+\left|\alpha\right|r_0^2+\left|\delta\right|}{3\left|\alpha\right|r_0}$, $s=\frac{\left|\delta\right|}{3\left|\alpha\right|}$.
 \end{table}
Further, the qualitative nature of the energy density for these three types of wormholes are presented graphically in figures (\ref{fig:f1})--(\ref{fig:f3}).
\\
 \section{Wormhole space-time geometry and Embedding}
 \label{sec:iii}
 Due to spherically symmetric nature and static character of the wormhole geometry, the two dimensional hypersurface $H$ : $t=$ constant, $\theta=\pi/2$ describes the wormhole topology  \cite{r20}, having line element
 \begin{equation} \label{eq25} d{S_{H}^2}=\frac{dr^2}{1-\frac{b(r)}{r}}+r^2d\phi^2=dR^2+r^2(R)d\phi^2.
  \end{equation}
  Also this hypersurface $H$ can be considered as embedded rotational surface $z=z(r, \phi)$ in the Euclidean $3D$ space  ($r$, $\phi$, $z$) so that one has
    \begin{equation} \label{eq26} d{S^2_{H}}=\left[1+\left(\frac{dz}{dr}\right)^2\right]dr^2+r^2d\phi^2 .\end{equation}
     Now comparing (\ref{eq25}) and (\ref{eq26}) gives expression for embedding shape
      function as 
  \begin{equation} \label{eq27} z(r)=\int_{r_0}^{r}\sqrt{\frac{(b/r)}{1-(b/r)}}dr=\int_{r_0}^{r}\sqrt{\frac{1-\left(1-\frac{r_0}{r}\right)\left(\frac{\delta}{r_0}-\alpha r\right)}{\left(1-\frac{r_0}{r}\right)\left(\frac{\delta}{r_0}-\alpha r\right)}}dr
  \end{equation}
  which shows that embedding is possible if the function $b(r)$ to be positive definite. From equation (\ref{eq10}), as $b(r)$ is a quadratic in $r$ so positivity of $b(r)$ depends on the sign of discriminant
  \begin{equation}\label{eq28}
  \Theta=\bigg[\bigl(\alpha r_0-\frac{\delta}{r_0}-1\bigr)^2-\frac{4\delta}{r_0}\bigg].
  \end{equation}
  In the following table various possibilities for embedding, range of `$r$' and restrictions on the parameters are shown:
  \begin{table}[!htb]
  	
  	\caption{Embedding conditions, range of $`r$' and restrictions}
  	\begin{tabular}{|>{\bfseries}c|*{7}{c|}}\hline
  		{Sign of $\Theta$} & {\bf Embedding } & {\bf Embedding} & {\bf Embedding} & {\bf Range of $r$} & {\bf Restriction on}\\ 
  		{} & {\bf for WH 1} & {\bf for WH 2} &{\bf for WH 3} & &{\bf the parameters}
  		\\ \hline
  		\text{$\Theta=0$}  & \text{possible} & {possible} & {impossible} & {($r_0$, $\infty$)} & {$\alpha r_0=\bigl(\sqrt{\frac{\delta}{r_0}}\pm1\bigr)^2$}     \\ \hline
  		\text{$\Theta<0$}  & \text{possible} & {possible} & {impossible} & {($r_0$, $\infty$)} & {$\alpha r_0>\bigl(\sqrt{\frac{\delta}{r_0}}\pm1\bigr)^2$}  \\	\hline
  		\text{$\Theta>0$}  & \text{possible} & {possible} & {possible with } & {($-\infty$, $r_1$)$ \bigcup $($r_2$, $\infty$)} & {$\alpha r_0>\bigl(\sqrt{\frac{\delta}{r_0}}+1\bigr)^2$ for WH 1,} \\
  		& & &{no restriction}& & {$\alpha r_0<\bigl(\sqrt{\frac{\delta}{r_0}}-1\bigr)^2$ for WH 2} \\ 
  		&&&{on the parameters}&&
  		\\ \hline
  		
  	\end{tabular}
  \label{Table:T5}\\
  ~~~~~~~~~~~~~~~~~~~~~~~~~~~~~~~~~~~~~~~~~~~~~~~~~~~~~~~~~~~~~~~~~~~~~~~~~~~~~~~~~~~~~~~~~~~~~~~~~~~~~~~~~~~~~~~~~~~~~~~~~~~~~~~~~~~~~~~~~~~~~~~~~~~~~~~~~~~~~~~~~~~~~~~~~~~~~~~~~~~~~~~~~~~~~~~~~~~~~~~~~~~~~~~~~~~~~~~~~~~~~~~~~~~~~~~~~~~~~~~~~~~~~~~~~~~~~~~~~~~~~~~~~~~~~~~~~~~~~~~~~~~~~~~~~~~~~~~~~~~~~~~~~~~~~~~~~~~~~~~~~~~~~~~~~~~~~~~~~~~~~~~~~~~~~~~~~~~~~~~~~~~~~~~~~~~~~~~~~~~~~~~~~ 
\text{,} $\text{where}~~r_1=\frac{-\bigl(\alpha r_0-\frac{\delta}{r_0}-1\bigr)-\sqrt{\Theta}}{2\alpha}$, $r_2=\frac{-\bigl(\alpha r_0-\frac{\delta}{r_0}-1\bigr)+\sqrt{\Theta}}{2\alpha}$ ~~ ($r_0<r_1$).
\end{table}\\
\par
\begin{figure}[!htb]
 \includegraphics[width=.4 \textwidth]{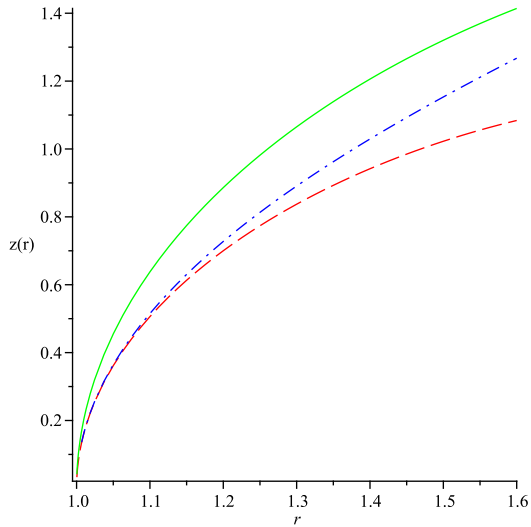}
 \caption{ Embedding for the function in (\ref{eq27}). For all plots the throat is located at $r_0$. The dash line describes the case ($\delta=0.5, \alpha=-1$). The dashdot and solid lines represent the cases ($\delta=2, \alpha=0.5$) and ($\delta=-0.05, \alpha=-1$) respectively.}
 \label{fig:f4}
 \end{figure}
\begin{figure}[!htb]
 	\centering
 	\begin{minipage}{.5\textwidth}
 		\centering
 		\includegraphics[width=.6\linewidth]{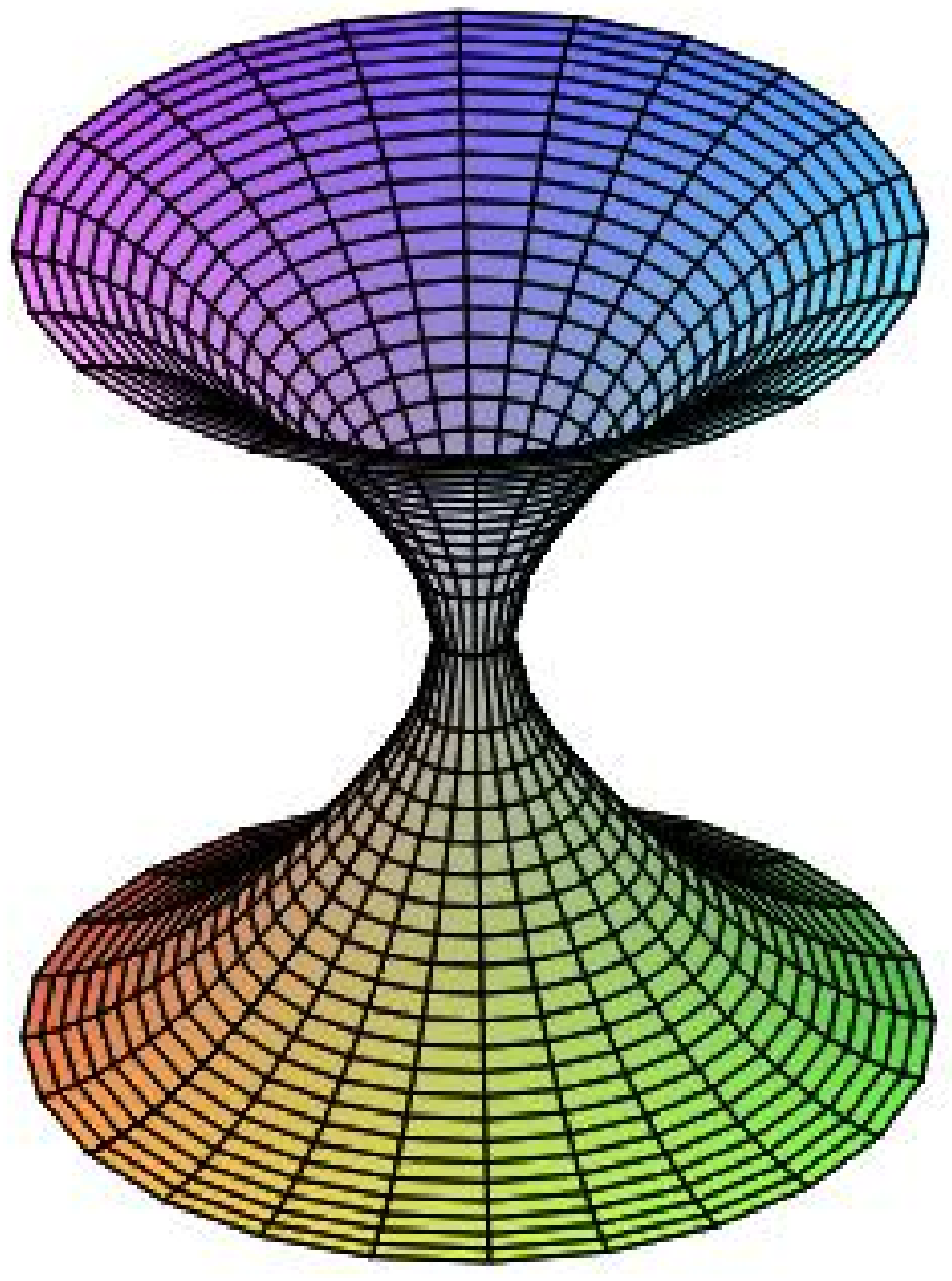}
 		\caption{Approximate embedding function for WH 1\\
 			 ($\alpha=0.005$, $\delta=0.4$) }
 		\label{fig:f5}
 	\end{minipage}
 	\begin{minipage}{.5\textwidth}
 		\centering
 		\includegraphics[width=.6\linewidth]{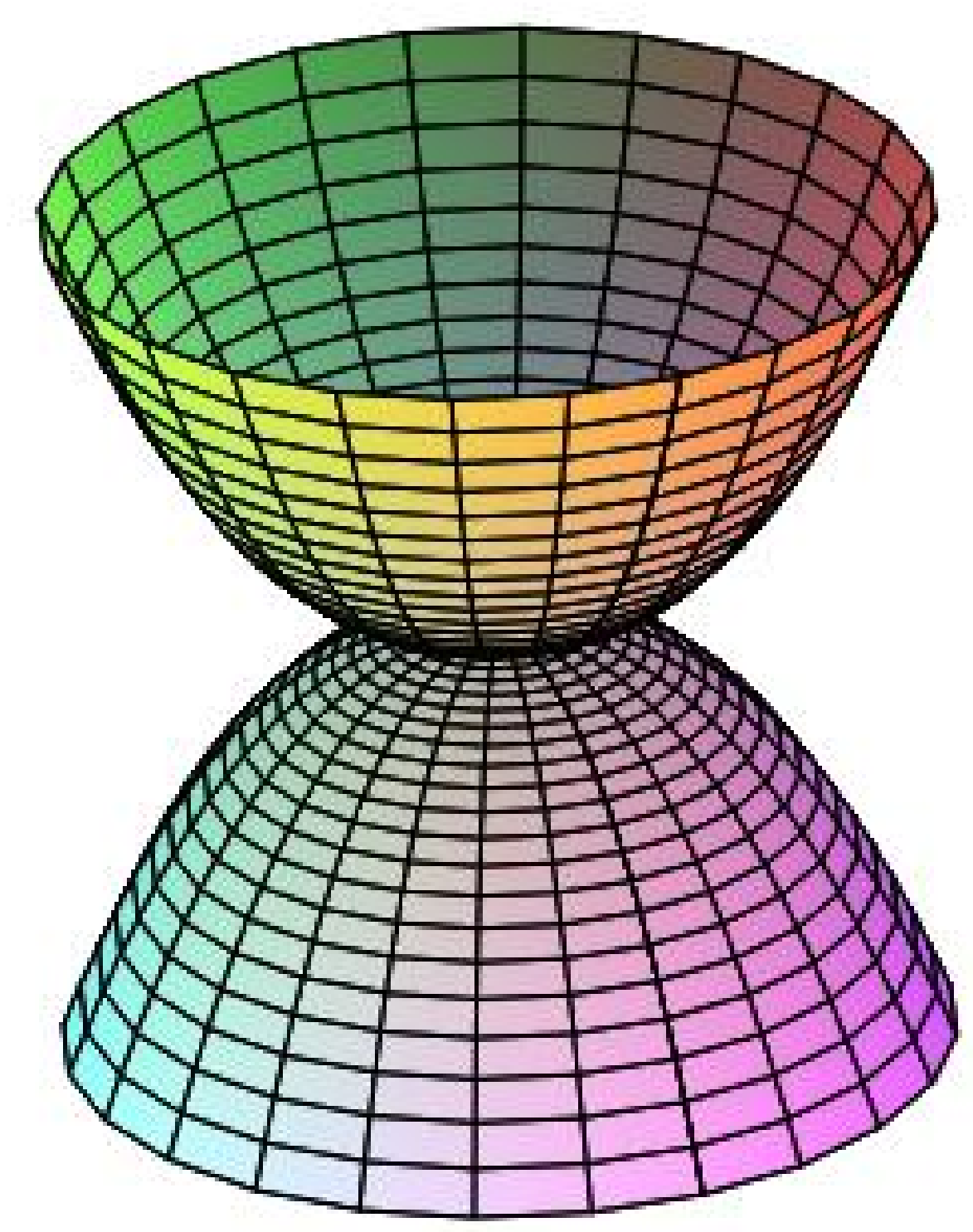}
 		\caption{ Approximate embedding function for WH 2\\ ($\alpha=-2$, $\delta=0.05$) }
 		\label{fig:f6}
 	\end{minipage}
 \end{figure}
\begin{figure}[!htb]
 	\includegraphics[width=.4 \textwidth]{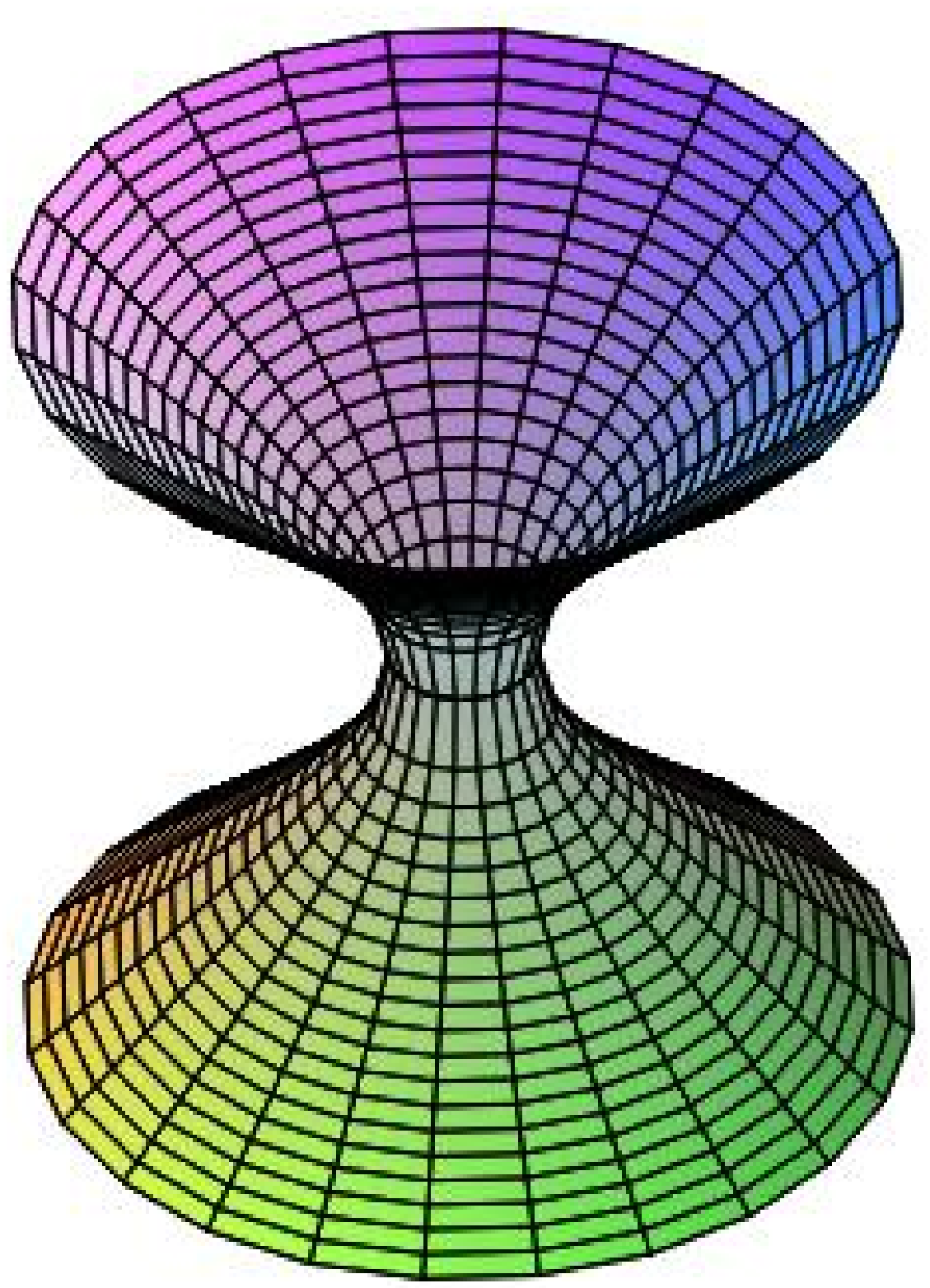}
 	\caption{ Approximate embedding function for WH 3 ($\alpha=-0.5$, $\delta=-0.04$) }
 	\label{fig:f7}
 \end{figure}

On the other hand, due to complicated expression of the integrand in equation(\ref{eq27}) it is not possible to have an analytic expression for $z(r)$. Figure (\ref{fig:f4}) shows the shape of the embedding function for three types of wormholes. However, it is possible to have an approximate expression of $z(r)$ near the throat as (for details see the appendix \ref{sec:emb})
\begin{equation}\label{eq29.1}
z(r)\approxeq\frac{2r_0\epsilon^{\frac{1}{2}}}{\sqrt{(\delta-\alpha r_0^2)}}\Bigg[1+\frac{A}{3}\epsilon+\frac{B}{5}\epsilon^2+O(\epsilon^3)\Bigg]
\end{equation}
with $\epsilon=r-r_0$, $A=\frac{1}{2r_0^2}\Biggl\{\frac{\alpha r_0^3}{\delta-\alpha r_0^2}+r_0+\alpha r_0^2-\delta\Biggr\}$ and $B=\frac{1}{2}\frac{\alpha\delta}{(\delta-\alpha r_0^2)^2}-\frac{1}{2}A^2$.\\
The graphical presentation of this approximate embedding function is shown in figures (\ref{fig:f5})--(\ref{fig:f7}) for the present 3 wormhole geometries.
 \\
\par 
 \section{ Geodesics in wormhole spacetime: Hamilton-Jacobi approach}\label{sec:iv}
 In classical mechanics, the geodesic equations describing the test particle motion in a given spacetime can be derived from the Hamilton-Jacobi equation \cite{r25}
  
 \begin{equation}\label{eq29}
 \frac{\partial S}{\partial \lambda}+\frac{1}{2}g^{\mu\nu}\frac{\partial S}{\partial x^\mu}\frac{\partial S}{\partial x^\nu}=0
 \end{equation} where $S$ is the usual Hamilton-Jacobi action, $g^{\mu\nu}$ are the metric components of the spacetime having coordinates $x^\mu=\{t,r,\theta,\phi\}$ and $\lambda$ is the affine parameter along the geodesic. As the metric tensor of the present wormhole geometry are independent of the temporal co-ordinate $`t$' and spatial co-ordinate $`\phi$' so wormhole space-time has killing vector fields along these two co-ordinate lines. More precisely the orbits of the space-like killing field are parametrized by $\phi$. Further, due to stationary and axisymmetric nature of the wormhole geometry there are two constants of motion associated with these killing vector field and are identified as the energy ($E$) and angular momentum ($L$) (about the axis of symmetry) of the test particle. Also, due to spherical symmetry of the space-time it is possible to choose co-ordinates so that the geodesic is always on the equatorial plane ($\theta=\pi/2$). As a consequence, the solution ansatz for the above Hamilton-Jacobi equation can be written in the following separable form \cite{r19}:
  \begin{equation}\label{eq30}
 S=-Et+L\phi+S_1(r)+\frac{1}{2}m^2\lambda ~,
 \end{equation}
  where $S_1(r)$ is a function of $r$ alone, and $m$ is the mass of the test particle.\\
 The explicit form of the above Hamilton-Jacobi equation(\ref{eq29}) for the the present wormhole spacetime is
  \begin{equation}\label{eq31}
 -\left(\frac{\partial S}{\partial t}\right)^2+\Bigg\{1-\frac{b(r)}{r}\Bigg\}\left(\frac{\partial S}{\partial r}\right)^2+\frac{1}{r^2}\left(\frac{\partial S}{\partial \phi}\right)^2+m^2=0.
 \end{equation}
 Using the ansatz (\ref{eq30}) in equation (\ref{eq31}) one has the following expression (in integral form) for the unknown function $S_1$ as 
 \begin{subequations}
 	\begin{equation}\label{eq32a}
 	\left(1-\frac{b}{r}\right)\left(\frac{dS_1}{dr}\right)^2=E^2-m^2-\frac{L^2}{r^2}
 	\end{equation} {\it i.e.,}
 	\begin{equation}\label{eq32b}
 	S_1(r)=\epsilon\int\frac{\Big[E^2-m^2-\frac{L^2}{r^2}\Big]^\frac{1}{2}}{\sqrt{\left(1-\frac{b}{r}\right)}}dr.
 	\end{equation}
 \end{subequations}
Here $\epsilon=\pm1$ indicates the change of sign whenever $r$ passes through a zero in the integrand (\ref{eq32b}) and $L$, the separation constant, is known as Carter constant \cite{r15}. Now using equation (\ref{eq32b}) in equation (\ref{eq30}) gives the explicit form of the Hamilton-Jacobi action as
 \begin{equation} \label{eq34}
S(r,\phi,t)=\frac{1}{2}m^2\lambda-Et+L\phi+\epsilon\int\sqrt{\frac{E^2-m^2-\frac{L^2}{r^2}}{1-\frac{b}{r}}}dr.
\end{equation}
 For the particle trajectory following the Hamilton-Jacobi method we have $\frac{\partial S}{\partial E}$=constant, $\frac{\partial S}{\partial L}$=constant. 
Now choosing the above constants to be zero without any loss of generality one obtains 
\begin{equation}\label{eq35}
t=\epsilon\int\frac{E}{\sqrt{\left(1-\frac{b}{r}\right)\left(E^2-m^2-\frac{L^2}{r^2}\right)}}dr~,
\end{equation}
and
\begin{equation}\label{eq36}
\phi=\epsilon\int\frac{L}{r^2\sqrt{\left(1-\frac{b}{r}\right)\left(E^2-m^2-\frac{L^2}{r^2}\right)}}dr.
\end{equation} 
\\
Thus the equations (\ref{eq35}) and (\ref{eq36}) describe the particle trajectory in the wormhole geometry with 3 physical parameters namely $E$, $L$ and $m$ of which only two are independent. Now equation (\ref{eq35}) gives the radial velocity of the particle as
\begin{equation}\label{eq38}
v_r=\frac{dr}{dt}=
\begin{cases}
\sqrt{\frac{\alpha}{r}(r-r_0)(\mu_0 r_0-r)\left(1-\mu^2-\frac{d^2}{r^2}\right)},\text{ for WH 1}\\
\sqrt{\frac{|\alpha|}{r}(r^2-r_0^2)\left(1-\mu^2-\frac{d^2}{r^2}\right)},\text{ for WH 2}\\
\sqrt{\frac{|\alpha|}{r}(r-r_0)^2\left(1-\mu^2-\frac{d^2}{r^2}\right)},\text{ for WH 3}

\end{cases}
\end{equation}
where $\mu=\frac{m}{E}$ and $d=\frac{L}{E}$, are the impact parameters and the turning points of the trajectory are given by $\frac{dr}{dt}=0$. As a consequence, the potential curves are
\begin{equation}
\mu=\left(1-\frac{d^2}{r^2}\right)^{\frac{1}{2}}.
\end{equation} 
Due to monotonic decreasing nature of the potential curve in the radial coordinate `$r$', no extrema point exists so that test particle smoothly goes across the wormhole. Moreover, the photon trajectories (with $m=0$) will be analyze further in section VI.
\par 
 On the other hand, one may evaluate the integral on the R.H.S of equations (\ref{eq35}) and (\ref{eq36}) in terms of incomplete elliptic functions, so that the explicit form of the particle trajectory are given by 
 \begin{equation}
 t=
 \begin{cases}
 \frac{\epsilon}{\sqrt{|\alpha|}\sqrt{1-\mu^2}}\big[Q(r)-Q(r_0)\big], \text{ for WH 2}\\
 \frac{\epsilon}{\sqrt{|\alpha|}\sqrt{1-\mu^2}}\big[W(r)-W(r_0)\big], \text{ for WH 3}
 \end{cases}
 \end{equation}
 where 
 \begin{equation*}
 Q(r)=\frac{18\zeta^2r^{\frac{5}{2}}r_0^2F_1(\frac{5}{4}, \frac{1}{2}, \frac{1}{2}, \frac{9}{4},\frac{r^2}{\zeta^2}, \frac{r^2}{r_0^2})}{5\sqrt{(-\zeta^2+r^2)(r^2-r_0^2)}\biggl[9\zeta^2r_0^2F_1(\frac{5}{4}, \frac{1}{2}, \frac{1}{2}, \frac{9}{4},\frac{r^2}{\zeta^2}, \frac{r^2}{r_0^2})+2r^2\bigg(\zeta^2F_1(\frac{9}{4}, \frac{1}{2}, \frac{3}{2}, \frac{13}{4},\frac{r^2}{\zeta^2}, \frac{r^2}{r_0^2})+r_0^2F_1(\frac{9}{4}, \frac{3}{2}, \frac{1}{2}, \frac{13}{4},\frac{r^2}{\zeta^2}, \frac{r^2}{r_0^2}) \bigg)\biggr]}
 \end{equation*}
  and
 \begin{eqnarray}
 W(r)&=&\frac{1}{\sqrt{r}(\zeta+r_0)(\zeta^2-r^2)}\Bigg[(r^2-\zeta^2)\sqrt{\frac{2r}{\zeta}}\biggl\{2\zeta^2E\biggl(\sqrt{\frac{\zeta+r}{\zeta}},\frac{1}{2}\sqrt{2}\biggr)-\zeta^2F\bigg(\sqrt{\frac{\zeta+r}{\zeta}},\frac{1}{2}\sqrt{2}\bigg)\\\nonumber
 &~&-2\zeta F\bigg(\sqrt{\frac{\zeta+r}{\zeta}},\frac{1}{2}\sqrt{2}\bigg)r_0+2\zeta E\bigg(\sqrt{\frac{\zeta+r}{\zeta}},\frac{1}{2}\sqrt{2}\bigg)r_0+r_0^2\Pi\bigg(\sqrt{\frac{\zeta+r}{\zeta}}, \frac{\zeta}{\zeta+r_0}, \frac{1}{2}\sqrt{2}\bigg)\\\nonumber
 &~&-F\bigg(\sqrt{\frac{\zeta+r}{\zeta}},\frac{1}{2}\sqrt{2}\bigg)r_0^2\biggr\}\Bigg],\\\nonumber
 \end{eqnarray}
 \begin{equation}
 \phi=
 \begin{cases}
 \frac{\epsilon d}{\sqrt{|\alpha|}\sqrt{1-\mu^2}}\big[Q_1(r)-Q_1(r_0)\big], \text{ for WH 2}\\
 \frac{\epsilon d}{\sqrt{|\alpha|}\sqrt{1-\mu^2}}\big[W_1(r)-W_1(r_0)\big], \text{ for WH 3}
 \end{cases}
 \end{equation}
 where
 \begin{equation*}
 Q_1(r)=\frac{2\sqrt{r}F_1(\frac{1}{4}, \frac{1}{2}, \frac{1}{2}, \frac{3}{4},\frac{r^2}{\zeta^2}, \frac{r^2}{r_0^2})}{r_0\zeta}
 \end{equation*}
 and 
 \begin{equation*}
 W_1(r)=-\frac{\sqrt{2}}{\sqrt{\zeta}(\zeta+r_0)}\Pi\bigg(\sqrt{\frac{\zeta+r}{\zeta}}, \frac{\zeta}{\zeta+r_0}, \frac{1}{2}\sqrt{2}\bigg)
 \end{equation*}
  with $\zeta^2=\frac{d^2}{1-\mu^2}$, where $F_1$ is the hypergeometric series and $F,~E,~\Pi$ are incomplete elliptic integrals of the first kind, second kind and third kind respectively.

 \section{Null geodesics: Photon spheres and shadow of wormhole}
\label{sec:V}
 For a free falling particle the momenta one-forms are related to the background geometry by the geodesic equation
\begin{equation}
\frac{dp_a}{d\lambda}=\frac{1}{2}g_{ab, c}p^bp^c
 \end{equation}
 with $`\lambda$' the affine parameter. It is well known from the above relation that the momenta along those co-ordinates will be constants of motion if they are absent in all the metric coefficients. So in the present wormhole geometry if we confine ourselves to the equatorial plane ({\it i.e.,} $\theta=\pi/2$) then $t$ and $\phi$ are cyclic co-ordinates with the constants of motion $p_t$ and $p_\phi$ set as 
 \begin{equation}
 p_t=+E~~ \text{and}~~p_\phi=L,
 \end{equation}
 identifying $E$ as the energy and $L$ as the angular momentum of the particle as measured by an observer at asymptotically flat region (far from the source). Thus , we have \begin{equation}\label{neq3}
 p^t=\dot{t}=e^{-2\phi(r)}E,~~p^\phi=\dot{\phi}=\frac{L}{r^2},~~p^r=\frac{dr}{d\lambda}=\dot{r}.
 \end{equation}
 For null geodesic, the norm of momentum one-form vanishes {\it i.e.,} $p^ap_a=0$. This implies 
 \begin{equation}\label{neq4}
 \dot{r}^2=\left(1-\frac{b(r)}{r}\right)\left[e^{-2\phi(r)}E^2-\frac{L^2}{r^2}\right].
 \end{equation}
 Using (\ref{neq3}) in (\ref{neq4}), the differential equation for photon trajectory takes the form
 \begin{equation}\label{neq5}
 \left(\frac{dr}{d\phi}\right)^2=\frac{r^4\left(1-\frac{b(r)}{r}\right)}{\mu^2}\left[e^{-2\phi(r)}-\frac{\mu^2}{r^2}\right],
 \end{equation} with $\mu=\frac{L}{E}$, the impact parameter.
 Now the photons coming from infinity will not reach the throat of the wormhole ($r=r_0$) if there exists a turning point ({\it i.e., }$\dot{r}=0$) $r_s>r_0$. Here $r_s$ is termed as the distance of closest approach and impact parameter is given by 
 \begin{equation}
 \mu=\pm r_se^{-\phi(r_s)}.
 \end{equation}
 So the differential equation for photon trajectories takes the form
 \begin{equation}
 \frac{d\phi}{dr}=\pm\sqrt{\frac{1}{r^2\left(1-\frac{b(r)}{r}\right)}\left[e^{\phi(r_0)-\phi(r)}.\frac{r^2}{r_0^2}-1\right]^{-1}}.
 \end{equation}
 Alternatively, photon might get trapped in a sphere of constant $r$ and consequently it will not be able to reach the final infinite point. Such spheres are termed as photon spheres. So a photon sphere is a region where the curvature of space-time forces even null geodesics to move in circles. We shall discuss it shortly.
 Also at the location of the photon sphere both $\dot{R}$ and $\ddot{R}$ vanish for a photon. Now from equation (\ref{neq4}) we have 
 \begin{equation}
 {\dot{R}}^2=e^{-2\phi(r)E^2}-\frac{L^2}{r^2}.
 \end{equation}
 So at the photon sphere$:\frac{L^2}{r^2}=r^2e^{-2\phi(r)}$. Also for $r>r_0$ the condition for existence of photon sphere is given by $\frac{e^{\phi(r)}}{r}=1$. 
 \par
 On the other hand, due to co-ordinate singularity at the throat({\it i.e.,$r=r_0$}) those geodesics which passes through the throat from one side to the other of the wormhole, can not be described by equation(\ref{neq5}). To resolve this issue, the radial co-ordinate should be replaced by the proper radial distance($R$) as 
 \begin{equation}
 R(r)=\pm\int_{r_0}^{r}\sqrt{\left(1-\frac{b(r^{'})}{r^{'}}\right)}dr^{'}.
 \end{equation}  
 So $R(r)$ is positive for $r>r_0$, negative for $r<r_0$ and is zero at the throat. The differential equation(\ref{neq5}) of null geodesic with proper radial distance takes the form 
 \begin{equation}\label{neq11}
 \left(\frac{dR}{d\phi}\right)^2=\frac{r^4}{\mu^2}\left[e^{-2\phi(r)}-\frac{\mu^2}{r^2}\right].
 \end{equation} 
Now we shall examine whether it is possible to have shadow of the wormholes \cite{r25}, \cite{r27} discussed above. So only WH 2 and WH 3 are considered as their geometry is extended upto infinity \text{\it i.e.,} these wormholes may be considered as shortcut path between two distant regions termed as region A and region B. Now suppose there is a light source in region A while B does not have any light source near the wormhole throat. So there are two possibilities for the light rays emitting from the source in region A --- either propagated into the wormhole through its throat or scattered away from the wormhole. Hence a distant observer in region A will only visualize the photons scattered away from the wormhole while those photons absorbed by the wormhole will appear as dark spot which is termed as the shadow of the wormhole. Now the radial velocity of a particle along the null geodesic is given by (from equation (\ref{eq38}) with $m=0$) 
 \begin{equation}\label{eq58}
 \frac{dr}{d\lambda}=\sqrt{\left(1-\frac{d^2}{r^2}\right)\left(1-\frac{b}{r}\right)}.
 \end{equation}
 Hence one may write it similar to energy equations
 \begin{equation}
 \left(\frac{dr}{d\lambda}\right)+V_0=1,
 \end{equation}
 with 
 \begin{equation}
 V_0=1-\left(1-\frac{b}{r}\right)\left(1-\frac{d^2}{r^2}\right),
 \end{equation}
 behaves as potential.
 \par
Now the characterization of the light trajectory\cite{r28} to be scattered away from the wormhole is determined by the turning point $\left(\frac{dr}{d\lambda}=0\right)$ of the radial motion. Thus, there should be a critical orbit separating the escape and plunge motion and it corresponds to the highest among the maximum of the above potential like term $V_0$. Due to radial dependence only of the radial velocity in equation(\ref{eq58}), the critical orbit will be a spherical path with unstable nature. Thus for critical orbit one has ($r>r_0$) 
\begin{equation}\label{eq58.1}
\begin{split}
&(i)~V_0=1,~~(ii)~ \frac{dV_0}{dr}=0,~~(iii)~ \frac{d^2V_0}{dr^2}\leq0.
\end{split}
\end{equation} 
For infinitely extended wormhole configurations for WH 2 and WH 3, the explicit expression for $V_0$ is given by
\begin{equation}\label{eq59.1}
	V_0= 
\begin{cases}
1-\bigl(1-\frac{r_0}{r}\bigr)u_2(r)\\
1-\bigl(1-\frac{r_0}{r}\bigr)^2u_3(r)
\end{cases}
\end{equation}	
with $u_2(r)=|\alpha|(r+r_0)\bigl(1-\frac{d^2}{r^2}\bigr)$ and $u_3(r)=|\alpha|\bigl(r-\frac{d^2}{r}\bigr)$. 
Hence the above restriction (\ref{eq58.1}) can be written as 
\begin{equation}\label{eq60.1}
\begin{split}
&(i)~u_i(r)=0,~(ii)~\frac{du_i}{dr}=0,~(iii)~\frac{d^2u_i}{dr^2}\geq0, ~i=2,3.
\end{split}
\end{equation}
These conditions(\ref{eq60.1}) for the above 2 wormholes are represented in the following table:
\begin{table}[h]
	\centering
	\caption{Conditions for formation of wormhole shadow}
	\begin{tabular}{|>{\bfseries}c|*{5}{c|}}\hline
		{\bfseries Wormhole} & {$u_i(r)$} & {\bfseries $u_i=0$} & {\bfseries $\frac{du_i}{dr}=0$} & {\bfseries $\frac{d^2u_i}{dr^2}$}
		\\ \hline
		\text{WH 2} & $u_2(r)=|\alpha|(r+r_0)\bigl(1-\frac{d^2}{r^2}\bigr)$ & $r=d$ & \text{$r^3+d^2r+2r_0d^2=0$} & $-2\frac{|\alpha|d^2}{r^3}\bigl(1+\frac{3r_0}{r}\bigr)<0$ \\
		& & &\text{having no  $+ve$ real root}&
		\\
		\hline
		\text{WH 3}& $u_3(r)=|\alpha|\bigl(r-\frac{d^2}{r}\bigr)$ & $r=d$ &\text{$1+\frac{d^2}{r^2}=0$}&$-2\frac{|\alpha|d^2}{r^3}<0$\\
		&  & &\text{having no real root}&
		\\ 
		\hline	
	\end{tabular}
\label{Table:T6}
\end{table}
\par 
Therefore, for the present wormhole geometry with quadratic shape function, it is not possible to have any critical orbit for the photon trajectories.
\section{Time-like geodesics: Bounded and Unbounded orbits}
\label{sec:vi}
For a time-like particle the momentum one form $p_\mu$ is a time-like vector having constant norm square as $p^\mu p_\mu=-m^2$, $m$ is the mass of the particle. Introducing $\tilde{E}=\frac{E}{m}$ and $\tilde{L}=\frac{L}{m}$ as the energy and angular momentum per unit mass, the above equation gives (as in null geodesic)
\begin{equation}\label{neq14}
\dot{r}^2=\left(1-\frac{b(r)}{r}\right)\left({\tilde{E}}^2e^{-2\phi(r)}-\frac{\tilde{L}^2}{r^2}-1\right).
\end{equation}
Thus any time-like particle always reaches the throat with zero radial velocity. Further, the differential equation for time-like trajectory can be written as (in terms of proper radial distance )
\begin{equation}
\left(\frac{dl}{d\phi}\right)^2=\frac{r^4}{\tilde{\mu}^2}\left[e^{-\phi(l)}-\frac{\tilde{\mu}^2}{r^2}-\frac{1}{\tilde{E}^2}\right]
\end{equation}
with $\tilde{\mu}=\frac{\tilde{L}}{\tilde{E}}$. Now differentiating (\ref{neq14}) once the second derivative of the radial co-ordinate becomes
\begin{equation}
\ddot{r}=\frac{-\left(\frac{b'(r)}{r}-\frac{b(r)}{r^2}\right)}{2(1-\frac{b(r)}{r})}\dot{r}^2-\frac{\left(1-\frac{b(r)}{r}\right)}{2}\left\{-2\tilde{E}^2e^{-2\phi(r)}.\phi'(r)+\frac{2\tilde{L}^2}{r^3}\right\}.
\end{equation} 
For a particle with zero initial velocity {\it i.e.,} $\dot{r}=0=\tilde{L}\ddot{r}\alpha-\phi'(r).$ Hence for zero tidal force wormhole, a time-like particle will remain at the initial position if there is no initial velocity. Also at the throat both $\dot{r}$ and $\ddot{r}$ vanish {\it i.e.,} a time-like particle reaches the throat with zero tidal velocity as well as zero radial acceleration. We shall now discuss both unbounded and bounded time-like orbits in two separate subsections.
\subsubsection{Unbounded orbits}
Suppose a time-like particle is falling from infinity and before reaching the throat it gets deflected at a radial distance $r_s$ then $r_s$ is a real solution of the equation ($\dot{r}^2=0$)
\begin{equation}
\tilde{E}^2e^{-2\phi(r_s)}-\frac{\tilde{L}^2}{r_s^2}=1.
\end{equation}
Using this relation the differential equation for time-like geodesic takes the form
\begin{equation}
\frac{d\phi}{dr}=\pm\frac{\sqrt{\left(1-\frac{b}{r}\right)}\left(\frac{\mu}{r^2}\right)}{\sqrt{\mu^2\left(\frac{1}{r_s^2}-\frac{1}{r^2}\right)+\left(e^{-2\phi(r)}-e^{2\phi(r_s)}\right)}}.
\end{equation} 
{\it i.e.,} the geodesic curve in integral fom reads as 
\begin{equation}
\phi=\pm\int_{r_0}^{\infty}\frac{\left(\frac{\mu}{r^2}\right)\sqrt{\left(1-\frac{b}{r}\right)}}{\sqrt{\mu^2\left(\frac{1}{r_s^2}-\frac{1}{r^2}\right)+\left(e^{-2\phi(r)}-e^{2\phi(r_s)}\right)}}dr.
\end{equation}
\subsubsection{Bounded orbits}
To study the bound time-like for the present wormhole geometry we assume non-vanishing tidal force in the form
\begin{equation}
e^{2\phi}=1-\frac{b(r)}{r}+\epsilon(r)
\end{equation} 
where $\epsilon(r)$ is a continuous function having significant effect near the throat, otherwise it is vanishingly small. However, $\epsilon(r)$ can not be identically zero due to traversable nature of the wormhole. So at far distance from the throat at the radial velocity equation(\ref{neq14}) simplifies to 
\begin{equation}
\dot{r}^2=\tilde{E}^2-V^2(r)
\end{equation} 
where 
\begin{equation}
V^2(r)=\left(1-\frac{b(r)}{r}\right)\left(\frac{\tilde{L}^2}{r^2}+1\right),
\end{equation}
is identified as an effective potential.
\par 
Thus the allowed region for a particle with energy $\tilde{E}$(at infinity) is characterized by $V(r)<\tilde{E}$ {\it i.e.,} bounded orbits are identified within these redii for which $V$ is less than its conserved energy $\tilde{E}$.
\par 
Further, any bound orbit in wormhole geometry can be either a circular orbit (stable or unstable) or an orbit oscillating around the radius of a stable circular orbit.
\par 
In order to have circular orbits both $\dot{r}$ and $\ddot{r}$ should vanish for at least some $r$. Now, 
\begin{equation}
\dot{r}=0\implies\tilde{E}=|{V(r)}|~{\text{and}}~\ddot{r}=0\implies\frac{\mathrm{d}}{\mathrm{d}r}V^2(r)=0.
\end{equation}
Thus circular orbits exist if the energy of the time-like particle coincides with an extremum of the effective potential. Further, the circular orbit is classified as stable one if the energy of the particle corresponds to a minimum of the potential while at the maximum or saddle point of the potential the circular orbit will be unstable in nature. For the present wormhole geometry the explicit form of the effective potential is 
\begin{equation}
V^2(r)=\left(1-\frac{r_0}{r}\right)\left(\frac{\delta}{r_0}-\alpha r\right)\left(\frac{\tilde{L}^2}{r^2}+1\right).
\end{equation}
For the above effective potential it is easy to see that there is at least one circular path for the time-like particle with radius as $r_0<r<\sqrt{3}r_0$ assuming $3\alpha r_0^2>\delta$ and the stability of the orbit can be characterized by the sign of the second derivative of the effective potential.
\section{Discussion and Concluding remarks}
\label{sec:vii}
The present work deals with static traversable wormholes with quadratic shape function. Due to wormhole conditions there are 3 possible types of wormhole geometry of which one (\text{\it i.e.} WH 1) is finitely extended while the other two connect two distant regions (or universes). For WH 1 the shape function is expressed as product of two linear factors (with distinct $+ve$ roots) while for WH 2 the distinct roots are equal in magnitude $(r_0)$ but opposite in sign and for WH3 there are two equal roots ($r_0$). The energy conditions are examined for each type of wormhole and are presented in tabular form (see tables \ref{Table:T2}--\ref{Table:T4}) as well as variation of the energy densities are shown graphically (see figures \ref{fig:f1}--\ref{fig:f3}). It is found that in some region of the wormhole geometry the energy conditions are violated. The possibility of embedding of the wormhole geometry to the Euclidean space has been investigated and results with restrictions on the parameters are shown in table \ref{Table:T5}. Also possible embedding near the throat has been approximately evaluated and the nature of the embedding function has been presented graphically in figures \ref{fig:f4}--\ref{fig:f7}.
\par 
The motivation for choosing quadratic shape function is as follow: in ref\cite{r20} quadratic shape function for the wormhole geometry has been used for the first time. They have obtained $(i)$ phantom wormhole spacetimes extending to infinity and $(ii)$ static spacetimes of finite size with phantom wormhole connected to an inhomogeneous and anisotropic spherically symmetric distribution of dark energy. In the 2nd wormhole configuration the phantom matter is confined to a finite region around throat and is connected to the dark energy distribution. One may note that the wormhole part does not obey the dominant energy condition but it is obeyed by the dark energy distribution.
\par 
The wormhole geometry with quadratic shape function is not asymptotically flat. So the matter distribution for wormholes extending to infinity must be cut off at some radius $r^*>r_0$ (larger than the throat radius) and connected to an exterior asymptotically flat space time (say vacuum Schwarzschild space time). Consequently, the traversability of the wormhole is done through the matching of the interior wormhole solution to the exterior Schwarzschild solution \cite{r29}-\cite{r30}.
\par 
Motion of test particle (time-like or null) has been studied in details in sections \ref{sec:iv}-\ref{sec:vi}. Hamiltonian-Jacobi approach has been employed to analyze time-like and null geodesics. For radial and circular geodesics it is possible to obtain simple analytic form of the geodesic paths, however for non-radial (\text{\it i.e.,} arbitrary) geodesics due to complicated form of the integrands the geodesic equations are mostly restricted in integral form. Finally, for photon trajectories, it is examined (in section \ref{sec:V}) whether one can obtain shadow of the wormhole or not. The restrictions for shadow formation are presented in table \ref{Table:T6} and it is found that no critical orbit is possible at any radial distance $r_c>r_0$. Therefore, for the present wormhole configurations with quadratic shape function it is not possible to form any shadow of the wormhole as normally one has for black hole configuration.

\section*{Acknowledgement}
Author BG acknowledges CSIR-UGC for awarding Junior Research Fellowship (1250/(CSIR-UGC NET JUNE 2019))
 and the Department of
Mathematics, Jadavpur University where a part of the work was completed. S.C. thanks Science and Engineering Research Board (SERB) for awarding
MATRICS Research Grant support (File No. MTR/2017/000407) and Inter University Center for
Astronomy and Astrophysics (IUCAA), Pune, India for their warm hospitality as a part of the work
was done during a visit.

\appendix
\section{Approximate Embedding function}
\label{sec:emb}
From equation (\ref{eq27}) we get
\begin{eqnarray}
z(r)&=&\int_{r_0}^{r}\sqrt{\frac{r+\frac{(r-r_0)(\alpha r_0r-\delta)}{r_0}}{\frac{(r-r_0)(\delta-\alpha r_0r)}{r_0}}}dr\\\nonumber
&=&\int_{r_0}^{r}\frac{\biggl\{rr_0+(r-r_0)(\alpha r_0r-\delta)\biggr\}^\frac{1}{2}}{\{(r-r_0)(\delta-\alpha r_0r)\}^\frac{1}{2}}dr
\end{eqnarray}
putting $r=r_0+\epsilon$,
\begin{eqnarray}
z(r)&=&\int_{\epsilon=0}^{\epsilon}\frac{[r_0^2+\epsilon(r_0+\alpha r_0^2-\delta)+\alpha r_0\epsilon^2]^\frac{1}{2}}{[\epsilon\{(\delta-\alpha r_0^2)-\alpha r_0\epsilon\}]^\frac{1}{2}}d\epsilon\\\nonumber
&=&\int_{\epsilon=0}^{\epsilon}\frac{1}{\sqrt{\epsilon}}\frac{[r_0^2+\epsilon(r_0+\alpha r_0^2-\delta)+\alpha r_0\epsilon^2]^\frac{1}{2}}{(\delta-\alpha r_0^2)^\frac{1}{2}}\biggl(1-\frac{\alpha r_0}{\delta-\alpha r_0^2}\epsilon\biggr)^{-\frac{1}{2}}d\epsilon\\\nonumber
&=&\int_{\epsilon=0}^{\epsilon}\frac{1}{\sqrt{\epsilon}}\frac{[r_0^2+\epsilon(r_0+\alpha r_0^2-\delta)+\alpha r_0\epsilon^2]^\frac{1}{2}}{(\delta-\alpha r_0^2)^\frac{1}{2}}\biggl[1+\frac{\alpha r_0}{\delta-\alpha r_0^2}\epsilon+\frac{\alpha^2 r_0^2}{(\delta-\alpha r_0^2)^2}\epsilon^2+O(\epsilon^3)\biggr]^{\frac{1}{2}}d\epsilon\\\nonumber
&=&\int_{\epsilon=0}^{\epsilon}\frac{1}{\sqrt{\epsilon}(\delta-\alpha r_0^2)^\frac{1}{2}}\biggl[r_0^2+\biggl(r_0+\alpha r_0^2-\delta+\frac{\alpha r_0^3}{\delta-\alpha r_0^2}\biggr)\epsilon+\frac{\alpha\delta r_0^2}{(\delta-\alpha r_0^2)^2}\epsilon^2+O(\epsilon^3)\biggr]^{\frac{1}{2}}d\epsilon\\\nonumber
&=&\frac{r_0}{(\delta-\alpha r_0^2)^\frac{1}{2}}\int_{\epsilon=0}^{\epsilon}\frac{1}{\sqrt{\epsilon}}\biggl[1+\frac{1}{r_0^2}\biggl(r_0+\alpha r_0^2-\delta+\frac{\alpha r_0^3}{\delta-\alpha r_0^2}\biggr)\epsilon+\frac{\alpha \delta}{(\delta-\alpha r_0^2)^2}\epsilon^2+O(\epsilon^3)\biggr]^{\frac{1}{2}}d\epsilon\\\nonumber
&=&\frac{r_0}{(\delta-\alpha r_0^2)^\frac{1}{2}}\int_{\epsilon=0}^{\epsilon}\biggl(\frac{1}{\sqrt{\epsilon}}+A\sqrt{\epsilon}+B\epsilon^\frac{3}{2}+O(\epsilon^3)\biggr)d\epsilon\\\nonumber
\end{eqnarray}
\begin{eqnarray}
&&=
\frac{2r_0}{(\delta-\alpha r_0^2)^\frac{1}{2}}\biggl(\sqrt{\epsilon}+\frac{A}{3}\epsilon^{\frac{3}{2}}+\frac{B}{5}\epsilon^\frac{5}{2}+O(\epsilon^3)\biggr)\\\nonumber\end{eqnarray}
where $A=\frac{1}{2r_0^2}\biggl(r_0+\alpha r_0^2-\delta+\frac{\alpha r_0^3}{\delta-\alpha r_0^2}\biggr)$ and $B=\frac{1}{2}\frac{\alpha \delta}{(\delta-\alpha r_0^2)^2}-\frac{1}{2}A^2$.
\end{document}